\documentclass[journal]{IEEEtran}

\usepackage[T1]{fontenc}
\usepackage{graphicx}
\usepackage[caption=false]{subfig}
\usepackage{mathrsfs,amsmath,amsfonts,amssymb,amsthm,amstext,amscd,amsxtra,amsopn,mathtools}
\usepackage{diagbox}
\usepackage{lettrine}
\usepackage{cite}
\usepackage{eucal}
\usepackage{soul}
\providecommand{\keywords}[1]{\textbf{\textit{Index terms---}} #1}
\usepackage{bm}
\usepackage{multirow}
\usepackage{esint}
\usepackage{empheq}
\usepackage{comment}
\usepackage{boxhandler}
\usepackage{algorithm}
\usepackage[noend]{algpseudocode}
\usepackage{etoolbox}
\apptocmd{\sloppy}{\hbadness 10000\relax}{}{}
\usepackage{leftidx}
\usepackage{makecell}
\usepackage{stackrel}
\usepackage{xcolor}
\usepackage{hyphenat}
\usepackage{nicefrac}
\usepackage{siunitx}
\usepackage[normalem]{ulem}
\usepackage[hyphens]{url}
\usepackage[hidelinks]{hyperref}
\usepackage{siunitx}
\hypersetup{breaklinks=true}
\newcommand{\matvec}[1]{\bm{#1}}

\algnewcommand{\IfThenElse}[3]{
  \State \algorithmicif\ #1\ \algorithmicthen\ #2\ \algorithmicelse\ #3}
\algnewcommand{\IfThen}[3]{
  \State \algorithmicif\ #1\ \algorithmicthen\ #2}

\DeclarePairedDelimiter\abs{\lvert}{\rvert}
\newcommand\norm[1]{\left\lVert#1\right\rVert}

\makeatletter
\def\BState{\State\hskip-\ALG@thistlm}

\makeatother

\algnewcommand{\Initialize}[1]{%
  \State \textbf{Initialize:}
  \Statex \hspace*{\algorithmicindent}\parbox[t]{.8\linewidth}{\raggedright #1}
}

\newcommand{\iu}{{\mathrm{i}}}

\let\oldref\ref
\renewcommand{\ref}[1]{(\oldref{#1})}

\makeatletter
\DeclareRobustCommand{\pder}[1]{%
  \@ifnextchar\bgroup{\@pder{#1}}{\@pder{}{#1}}}
\newcommand{\@pder}[2]{\frac{\partial#1}{\partial#2}}
\makeatother

\newcolumntype{M}[1]{>{\hbox to #1\bgroup\hss$}l<{$\egroup}}

\makeatletter
\newcommand\@brcolwidth{0.67em}

\def\@brarray[#1]{\array{r*\c@MaxMatrixCols {M{#1}}}}
\makeatother

\title{Global Maxwell Tomography Using the Volume-Surface Integral Equation for Improved Estimation of Electrical Properties}

\author{Ilias I. Giannakopoulos ~\IEEEmembership{Member,~IEEE}, Jos{\'e} E. Cruz Serrall{\'e}s ~\IEEEmembership{Member,~IEEE}, Jan Pa\v{s}ka, \\ Martijn A. Cloos, Ryan Brown, Riccardo Lattanzi, ~\IEEEmembership{Senior Member,~IEEE}

\thanks{This work was supported in part by NIH K99 EB035163 and NIH R01 EB024536, and was performed under the rubric of the Center for Advanced Imaging Innovation and Research (CAI$^2$R, www.cai2r.net), an NIBIB National Center for Biomedical Imaging and Bioengineering (NIH P41 EB017183). \textit{(Corresponding author: Ilias I. Giannakopoulos.)}}
\thanks{Ilias I. Giannakopoulos, Jos{\'e} E. Cruz Serrall{\'e}s, Ryan Brown, and Riccardo Lattanzi are with the Bernard and Irene Schwartz Center for Biomedical Imaging, Department of Radiology, New York University Grossman School of Medicine, NY, USA.}
\thanks{Ryan Brown and Riccardo Lattanzi are also with the Center for Advanced Imaging Innovation and Research (CAI$^2$R), Department of Radiology, New York University Grossman School of Medicine, NY, USA.}
\thanks{Jan Pa\v{s}ka is with the NeuroPoly Lab, Institute of Biomedical Engineering, Polytechnique Montr{\'e}al, Montreal, QC, Canada}
\thanks{Martijn A. Cloos is with the Donders Institute for Brain Cognition and Behaviour, Radboud University, Nijmegen, Netherlands and the Australian Institute for Bioengineering and Nanotechnology, University of Queensland, St Lucia, Queensland, Australia.}}

\begin{document}
\bstctlcite{IEEEexample:BSTcontrol}

\maketitle

\begin{abstract}
\textit{Objective:} Global Maxwell Tomography (GMT) is a noninvasive inverse optimization method for the estimation of electrical properties (EP) from magnetic resonance (MR) measurements. GMT uses the volume integral equation (VIE) in the forward problem and assumes that the sample has negligible effect on the coil currents. Consequently, GMT calculates the coil's incident fields with an initial EP distribution and keeps them constant for all optimization iterations. This can lead to erroneous reconstructions. This work introduces a novel version of GMT that replaces VIE with the volume-surface integral equation (VSIE), which recalculates the coil currents at every iteration based on updated EP estimates before computing the associated fields. \textit{Methods:} We simulated an 8-channel transceiver coil array for 7 T brain imaging and reconstructed the EP of a realistic head model using VSIE-based GMT. We built the coil, collected experimental MR measurements, and reconstructed EP of a two-compartment phantom. \textit{Results:} {\color{black}In simulations, VSIE-based GMT outperformed VIE-based GMT by at least 12\% for both EP}. In experiments, the relative difference with respect to probe-measured EP values in the inner (outer) compartment was 13\% (26\%) and 17\% (33\%) for the permittivity and conductivity, respectively. \textit{Conclusion:} The use of VSIE over VIE enhances GMT's performance by accounting for the effect of the EP on the coil currents. \textit{Significance:} VSIE-based GMT does not rely on an initial EP estimate, rendering it more suitable for experimental reconstructions compared to the VIE-based GMT.
\end{abstract}


\vspace{1cm}
\section*{Nomenclature}
\begin{IEEEdescription}[\IEEEsetlabelwidth{Notation }]
\item[Notation] Description      
\item[$a$] Scalar                                         		   
\item[$\matvec{a}$] Vector in $\mathbb{C}^{n}$           
\item[$A$] Matrix in $\mathbb{C}^{n_1 \times n_2}$   
\item[$\overline{(\cdot)}$] Conjugate         
\item[$(\cdot)^{\rm T}$] Transpose            
\item[$(\cdot)^{*}$] Hermitian adjoint      
\item[$\iu$] Imaginary Unit $\iu^2 = -1$ 
\item[$\circ$] Element\hyp wise Product
\item[$\otimes$] Kronecker Product
\item[{$[\dots]$}] Matrix Concatenation
\item[$\matvec{B}_1^{(+)}$] Vector of the Magnetic Transmit Field in $\mathbb{C}^{n}$
\end{IEEEdescription}

\section{Introduction} \label{sc:I}

\IEEEPARstart{T}{he} interactions of electromagnetic (EM) waves with biological tissue are determined by the underlying electrical properties (EP) of tissue \cite{hand2008modelling, collins2011calculation}. Access to EP at MRI frequencies could enable the estimation of local specific absorption rate maps \cite{ma2009towards, voigt2012patient}, which are important for ultra\hyp high field MRI \cite{lattanzi2009electrodynamic}, the accurate planning of radiofrequency (RF) and thermal\hyp based treatments \cite{rossmann2014review, hall2015cell, balidemaj2016hyperthermia}, and could lead to novel biomarkers for pathologies and treatment monitoring \cite{holder1992detection, ivorra2009vivo, balidemaj2015feasibility, shin2015initial}. Several techniques have been proposed to estimate EP from MR\hyp measurements by inverting the differential form of the Helmholtz wave equation \cite{zhang2014magnetic, liu2017electrical, hafalir2014convection}. Although fast, these methods result in artifacts associated with taking numerical derivatives of noisy data. Techniques based on the integral form of Maxwell's equations \cite{serralles2019noninvasive, giannakopoulos2020magnetic, balidemaj2015csi} are immune to these artifacts because the integration averages out additive noise. Despite showing superior accuracy in simulation, their experimental validation has been challenging because they require knowledge of the RF fields generated by the transmit coils (i.e., incident fields) to solve the forward problem. Given that the precise estimation of these incident fields is impractical, simplifying assumptions are usually made for their calculation \cite{serralles2019noninvasive,arduino2017csi}.
\par
Global Maxwell Tomography (GMT) is a technique for EP mapping based on the integral form of Maxwell's equations. GMT solves an iterative inverse problem to reconstruct the relative permittivity and electric conductivity of an object, using measurements of the transmit magnetic field \cite{serralles2019noninvasive, giannakopoulos2019global}. GMT relies on the volume integral equation (VIE) method for the forward problem \cite{georgakis2020fast} and on an ad-hoc regularization approach to mitigate noise in the measurements \cite{serralles2017investigation}. GMT demonstrated excellent performance in simulations with realistic anatomies \cite{giannakopoulos2020magnetic} and in experiments for homogeneous phantoms \cite{serralles2019noninvasive}. Its application to heterogeneous samples has been limited by the above-mentioned required knowledge of the incident fields. Specifically, in GMT, the incident fields are calculated once for an estimated EP distribution and then used throughout the optimization, assuming that the currents on the coil conductors are not affected by the updated guesses of EP at each iteration. This has prevented the application of GMT beyond proof of principle experiments with homogeneous phantoms. 
\par
To address this limitation, here we present a new implementation of GMT that uses the volume\hyp surface integral equation (VSIE) method \cite{Villena2016}. The proposed VSIE-based GMT does not assume knowledge of the incident fields, but calculates the coil currents at every iteration considering the latest estimate of EP and uses them to compute updated transmit magnetic fields at each iteration. A preliminary version of this work was presented in \cite{giannakopoulosnovel}. For the remainder of this work, we adhere to the notation defined in the nomenclature.

\section{Technical Background}

\subsection{Integral Equation Method}

\subsubsection{Volume Integral Equation}

The interactions between the electromagnetic fields and biological tissue can be modeled using the current\hyp based VIE as in \cite{Polimeridis2014, Oijala2014}. Precisely, the tissue model can be discretized over a voxelized grid, and the VIE can be solved for the unknown body currents using the Galerkin method of moments \cite{harrington1993field} by inverting a linear system:
\begin{equation}
Z_{\rm bb}\!\left(\matvec{\epsilon}\right) \matvec{j}_{\rm b}\!\left(\matvec{\epsilon}\right) = \text{RHS}(\matvec{e}_{\rm inc}).
\label{eq:vie}
\end{equation}
$Z_{\rm bb} \in \mathbb{C}^{(q \cdot n_v) \times (q \cdot n_v)}$ relates the induced polarization body current vector $\matvec{j}_{\rm b} \in \mathbb{C}^{(q \cdot n_v) \times 1}$ to the right hand side (RHS) of the VIE. The RHS is a function of the incident electric field $\matvec{e}_{\rm inc} \in \mathbb{C}^{(q \cdot n_v) \times 1}$ vector \cite{Polimeridis2014} (generated from an RF coil in MRI applications). $n_v$ is the number of the voxels of the discretization and $q$ are the basis function components per voxel. $\matvec{\epsilon} \in \mathbb{C}^{n_v \times 1}$ is the vector of the complex relative permittivities for all voxels, defined as:
\begin{equation}
\matvec{\epsilon} \equiv \matvec{\epsilon_r} + \frac{\matvec{\sigma_e}}{\iu \omega \epsilon_0},
\end{equation} 
where $\omega$ is the angular operating frequency of the MR scanner, $\epsilon_0$ is the permittivity of vacuum, and $\matvec{\epsilon_r}, \matvec{\sigma_e} \in \mathbb{R}^{n_v \times 1}$ contain the scalar relative permittivity and conductivity of the sample, respectively. Access to the body currents allows the estimation of the magnetic scattered field vector $\matvec{h}_{\rm sca} = K_{\rm bb}\matvec{j}_{\rm b} \in \mathbb{C}^{(q \cdot n_v) \times 1}$ with the aid of the discretized Green's function operator $K_{\rm bb} \in \mathbb{C}^{(q \cdot n_v) \times (q \cdot n_v)}$ \cite{Polimeridis2014}. The $\matvec{B}_1^{(+)} \in \mathbb{C}^{n_v \times 1}$ in the body can be expressed as:
\begin{equation}
\begin{aligned}
\matvec{B}_1^{(+)}\!\left(\matvec{\epsilon}\right) &= \mu_0 F \left( \matvec{h}_{\rm inc} + K_{bb}\matvec{j}_{\rm b}\!\left(\matvec{\epsilon}\right) \right), \\
F &=  G^{-1} \left([1 \:\: \iu \:\: 0] \otimes [1\:\: 0\:\: 0\:\: 0] \otimes I\right).
\end{aligned}
\label{eq:b1p_vie}
\end{equation}
Here, $\mu_0$ is the magnetic permeability of vacuum, $G$ is the Gram matrix, $\matvec{h}_{\rm inc} \in \mathbb{C}^{(q \cdot n_v) \times 1}$ is the incident magnetic field vector and $I \in \mathbb{N}^{n_v \times n_v}$ is the identity matrix.

\subsubsection{Volume-Surface Integral Equation}

One can omit the coil's incident fields ($\matvec{e}_{\rm inc}, \matvec{h}_{\rm inc}$) by including the coil conductors in the simulation domain and solving for both the body and the coil currents simultaneously. In this case, Eq. \ref{eq:vie} is replaced by the VSIE system \cite{Villena2016}:
\begin{equation}
\begin{bmatrix}
Z_{\rm cc} & Z_{\rm cb}^{\rm T} \\
Z_{\rm cb} & Z_{\rm bb}\!\left(\matvec{\epsilon}\right)       
\end{bmatrix} \begin{bmatrix}
\matvec{j}_{\rm c}\!\left(\matvec{\epsilon}\right) \\
\matvec{j}_{\rm b}\!\left(\matvec{\epsilon}\right) 
\end{bmatrix} = \begin{bmatrix}
\matvec{v} \\
\matvec{0}
\end{bmatrix}.
\label{eq:vsie}
\end{equation}
VSIE is a simplified domain decomposition method, where the Galerkin method is used to discretize the coil and the body currents. In particular, the coil conductors can be discretized over a triangular mesh with $m$ discretization elements and the coil current vector $\matvec{j}_{\rm c} \in \mathbb{C}^{m \times 1}$ can be approximated using the Rao\hyp Wilton\hyp Glisson basis functions \cite{Rao1982}. $Z_{\rm cc} \in \mathbb{C}^{m \times m}$ and $Z_{\rm cb} \in \mathbb{C}^{(q \cdot n_v) \times m}$ relate the coil currents with the induced voltage $\matvec{v} \in \mathbb{C}^{m \times 1}$ and the incident electric field, respectively. The $\matvec{B}_1^{(+)}$ can be computed as 
\begin{equation}
\matvec{B}_1^{(+)}\!\left(\matvec{\epsilon}\right) = \mu_0 F G^{-1} \left( K_{\rm cb}\matvec{j}_{\rm c}\!\left(\matvec{\epsilon}\right) + K_{\rm bb}\matvec{j}_{\rm b}\!\left(\matvec{\epsilon}\right) \right).
\label{eq:b1p_vsie}
\end{equation}
Here, $K_{\rm cb} \in \mathbb{C}^{(q \cdot n_v) \times m}$ relates the coil current vector with the incident magnetic field in each voxel of the sample \cite{Villena2016}.

\subsection{Global Maxwell Tomography}

The reconstruction of the EP with GMT is performed iteratively, using the quasi\hyp Newton optimization algorithm L\hyp BFGS\hyp B \cite{zhu1997algorithm}. In each iteration of the optimizer, the VIE system is solved using the EP guess from the previous iteration to generate synthetic $\matvec{B}_1^{(+)}$ maps using \eqref{eq:b1p_vie}. These maps are compared to the corresponding measured $\hat{\matvec{B}}_1^{(+)}$ maps within a cost function. This VIE\hyp based GMT algorithm requires a set of incident fields ($\matvec{e}_{\rm inc}$, $\matvec{h}_{\rm inc}$) as the input, which remains constant for all iterations. The cost function $f\!\left(\matvec{\epsilon}\right) = f_d\!\left(\matvec{\epsilon}\right) + \alpha f_r\!\left(\matvec{\epsilon}\right)$ that is minimized is composed by a weighted data\hyp consistency term $f_d$ and a regularization term $f_r$. $f_d$ is given from:
\begin{equation}
\begin{aligned}
f_d(\matvec{\epsilon}) &= \eta^{-1}\sqrt{\norm{w_l\circ w_{l'}\circ \matvec{\delta}_{ll'}\!\left(\matvec{\epsilon}\right)}_2^2}, \\
\eta &= \sqrt{\norm{w_l\circ w_{l'}\circ \hat{\matvec{B}}^{(+)}_{1,l} \circ \overline{\hat{\matvec{B}}^{(+)}_{1,l'}}}_2^2}, \\
\matvec{\delta}_{ll'}\!\left(\matvec{\epsilon}\right) &= \hat{\matvec{B}}^{(+)}_{1,l} \circ \overline{\hat{\matvec{B}}^{(+)}_{1,l'}} - \matvec{B}^{(+)}_{1,l}\!\left(\matvec{\epsilon}\right) \circ \overline{\matvec{B}^{(+)}_{1,l'}\!\left(\matvec{\epsilon}\right)},
\end{aligned}
\label{eq:cost_function}
\end{equation}
where Einstein summation notation is used over the indices $l'$ (first) and $l$ (second) that represent the number of measured $\hat{\matvec{B}}_1^{(+)}$ maps. $f_r(\matvec{\epsilon})$ is the so\hyp called Match Regularizer \cite{serralles2019noninvasive}, which smooths the effect of the noise in the measurements while preserving tissue boundaries. The parameter $\alpha \in \mathbb{R}$ is the weight of the regularizer. The weights $w \in \mathbb{R}^{n_v \times 1}$ in the data\hyp consistency term can be set to prioritize regions of the $\hat{\matvec{B}}_1^{(+)}$ maps with higher signal-to-noise ratio (SNR).
\par
The gradient of the data\hyp consistency term with respect to $\matvec{\epsilon}$ can be computed analytically through the solution of the Hermitian adjoint system of \eqref{eq:vie}. Specifically, one needs to first compute the derivative of $\matvec{B}_1^{(+)}$ with respect to $\matvec{\epsilon}$ using equation \eqref{eq:b1p_vie}: 
\begin{equation}
\begin{aligned}
\pder{\matvec{B}_1^{(+)}\!\left(\matvec{\epsilon}\right)}{\matvec{\epsilon}} &= \mu_0 F K_{bb} \pder{\matvec{j}_{\rm b}\!\left(\matvec{\epsilon}\right)}{\matvec{\epsilon}}.
\end{aligned}
\end{equation}
{\color{black} Note that the derivative of $\matvec{h}_{\rm inc}$ with respect to $\matvec{\epsilon}$ is zero}. The derivative of $\matvec{j}_{\rm b}$ with respect to $\matvec{\epsilon}$ requires the differentiation of the VIE \eqref{eq:vie} system as shown in \cite{serralles2019noninvasive}.

\section{Theory}

In this work, we propose a VSIE\hyp based implementation of GMT where the coil's incident EM fields are not calculated once and passed as input to the algorithm, but are rather implicitly re\hyp calculated for each updated guess of EP at every GMT iteration, by solving the VSIE for both coil currents and body currents. While the cost function and the optimization pipeline can remain the same as in the VIE\hyp based GMT, the synthetic $\matvec{B}_1^{(+)}$ maps must be computed using equation \eqref{eq:b1p_vsie}. Their derivatives with respect to $\matvec{\epsilon}$ are updated as follows:
\begin{equation}
\begin{aligned}
\pder{\matvec{B}_1^{(+)}\!\left(\matvec{\epsilon}\right)}{\matvec{\epsilon}} &= \mu_0 F \left( K_{bc}\pder{\matvec{j}_{\rm c}\!\left(\matvec{\epsilon}\right)}{\matvec{\epsilon}} + K_{bb}\pder{\matvec{j}_{\rm b}\!\left(\matvec{\epsilon}\right)}{\matvec{\epsilon}} \right), \:\: \text{or}\\
\pder{\matvec{B}_1^{(+)}\!\left(\matvec{\epsilon}\right)}{\matvec{\epsilon}} &= \mu_0 F \begin{bmatrix}
K_{bc} \\
K_{bb}
\end{bmatrix}^{\rm T}\pder{\matvec{j}_{\rm cb}\!\left(\matvec{\epsilon}\right)}{\matvec{\epsilon}}.
\end{aligned}
\label{eq:db1pder}
\end{equation}
Here, $\matvec{j}_{\rm cb}$ is the concatenated vector of both coil and body currents, and its derivative with respect to $\matvec{\epsilon}$ can be computed through the differentiation of the VSIE system \eqref{eq:vsie}. 
\par
The gradient of the data\hyp consistency term can be computed using Wirtinger calculus \cite{kreutz2009complex} (real function of complex arguments) as follows:
\begin{equation}
\begin{aligned}
\pder{f_d\!\left(\matvec{\epsilon}\right)}{\matvec{\epsilon}} &= \frac{-\left( \pder{\matvec{B}^{(+)}_{1,l}\!\left(\matvec{\epsilon}\right)}{\matvec{\epsilon}}\matvec{t}_{l,l'}\!\left(\matvec{\epsilon}\right) + \pder{\overline{\matvec{B}^{(+)}_{1,l}\!\left(\matvec{\epsilon}\right)}}{\matvec{\epsilon}} \overline{\matvec{t}_{l,l'}\!\left(\matvec{\epsilon}\right)} \right)}{\eta^2 f_d\!\left(\matvec{\epsilon}\right)}, \\
\matvec{t}_{l,l'}\!\left(\matvec{\epsilon}\right) &= \abs{\matvec{w}_l}^2 \circ \abs{\matvec{w}_{l'}}^2 \circ \overline{\matvec{\delta}_{ll'}\!\left(\matvec{\epsilon}\right)} \circ \overline{\matvec{B}^{(+)}_{1,l'}\!\left(\matvec{\epsilon}\right)}.
\end{aligned}
\label{eq:grad}
\end{equation}
Here, we used Einstein summation notation over the indices $l'$ (first) and $l$ (second). If RF shimming \cite{lattanzi2009electrodynamic} is not used to modulate the phase of the transmit coils, the derivative of the conjugate $\matvec{B}_1^{(+)}$ is zero. Otherwise, the conjugate derivative should also be computed, since the shimmed $\matvec{B}_1^{(+)}$ is a function of the conjugate unshimmed field. For example, let us assume that the $\matvec{B}_1^{(+)}$ is shimmed in such a way that the resulting phase at voxel $v$ is zero. The phase\hyp modulated ($\matvec{\beta}$) can be re\hyp written as:
\begin{equation}
\matvec{\beta}\!\left(\matvec{\epsilon}\right) = \matvec{B}_1^{(+)}\!\left(\matvec{\epsilon}\right) e^{-\iu \phi\!\left(\matvec{\epsilon}\right)} = \matvec{B}_1^{(+)}\!\left(\matvec{\epsilon}\right) \frac{\overline{\matvec{B}_{1,v}^{(+)}\!\left(\matvec{\epsilon}\right)}}{\abs{\matvec{B}_{1,v}^{(+)}\!\left(\matvec{\epsilon}\right)}}.
\label{eq:shimming}
\end{equation}
Here, $\phi$ is the phase modulation applied to $\matvec{B}_1^{(+)}$ to cancel the phase at voxel $v$. The derivatives that must be used in \eqref{eq:grad} can be computed as follows:
\begin{equation}
\begin{aligned}
\pder{\matvec{\beta}\!\left(\matvec{\epsilon}\right)}{\matvec{\epsilon}} &= e^{-\iu \phi\!\left(\matvec{\epsilon}\right)} \left(I-\frac{e^{-\iu \phi\!\left(\matvec{\epsilon}\right)} \matvec{B}_1^{(+)}\!\left(\matvec{\epsilon}\right)}{2\abs{\matvec{B}_{1,v}^{(+)}\!\left(\matvec{\epsilon}\right)}} I_{:v} \right)\pder{\matvec{B}_1^{(+)}\!\left(\matvec{\epsilon}\right)}{\matvec{\epsilon}} \\
\pder{\overline{\matvec{\beta}\!\left(\matvec{\epsilon}\right)}}{\matvec{\epsilon}} &= e^{-\iu \phi\!\left(\matvec{\epsilon}\right)} \frac{\overline{\matvec{B}_1^{(+)}\!\left(\matvec{\epsilon}\right)}}{2\abs{\matvec{B}_{1,v}^{(+)}\!\left(\matvec{\epsilon}\right)}} I_{:v} \pder{\matvec{B}_1^{(+)}\!\left(\matvec{\epsilon}\right)}{\matvec{\epsilon}},
\end{aligned}
\end{equation}
where $I_{:v}$ is the $v$\hyp th column of the identity matrix.
\par
{\color{black}Assuming that RF shimming is not used, the co\hyp gradient of the data\hyp consistency term can be written with the help of \eqref{eq:grad} as follows:
\begin{equation}
\left( \pder{f_d(\matvec{\epsilon})}{\matvec{\epsilon}} \right)^* = -\frac{\mu_0 X^T}{\eta^2 f_d(\matvec{\epsilon})} \left( \overline{\matvec{a}(\matvec{\epsilon})} \circ \matvec{\zeta}_l(\matvec{\epsilon})\right)_{l},
\end{equation}
where $X$ is
\begin{equation}
X = \begin{bmatrix}
    0^{m \times n_v} \\
    \matvec{1}_q \otimes I^{n_v \times n_v}
\end{bmatrix}, 
\end{equation}
and $\matvec{\alpha}$ is defined with the help of \eqref{eq:vsie} as
\begin{equation}
\matvec{a}(\matvec{\epsilon}) = - \pder{\begin{bmatrix}
Z_{\rm cc} & Z_{\rm cb}^{\rm T} \\
Z_{\rm cb} & Z_{\rm bb}\!\left(\matvec{\epsilon}\right)
\end{bmatrix}}{\matvec{\epsilon}}
\begin{bmatrix}
\matvec{j}_{\rm c}(\matvec{\epsilon}) \\
\matvec{j}_{\rm b}(\matvec{\epsilon})
\end{bmatrix}.
\label{eq:der_system}
\end{equation}
$\matvec{\zeta}_l$ can be computed via \eqref{eq:db1pder} as the solution of the Hermitian adjoint system of \eqref{eq:vsie} as follows:
\begin{equation}
\left(\begin{bmatrix}
Z_{\rm cc} & Z_{\rm cb}^{\rm T} \\
Z_{\rm cb} & Z_{\rm bb}\!\left(\matvec{\epsilon}\right)
\end{bmatrix}\right)^* \matvec{\zeta_l}(\matvec{\epsilon}) = \begin{bmatrix}
\overline{K_{bc}} \\
K_{bb}^*
\end{bmatrix} F^* \overline{\matvec{t}_{l,l'}\!\left(\matvec{\epsilon}\right)}
\end{equation}
Note that in \eqref{eq:der_system}, $\matvec{j}_{\rm c}(\matvec{\epsilon})$ and $\matvec{j}_{\rm b}(\matvec{\epsilon})$ have been computed by solving \eqref{eq:vsie} with the current guess of the object's EP.}

\section{Methods}\label{sec:Methods}

\subsection{Phantoms}\label{sec:phantoms}

\subsubsection{Numerical phantoms}\label{sec:phantoms_simulated}

We modeled four cylindrical phantoms with $\epsilon_r = 80$, $\sigma_e = \SI{0.6}{\siemens\per\meter}$, length \SI{17.2}{\centi\meter}
, and discretized them with \SI{5}{\milli\meter} voxel isotropic resolution. The cylinders had different radii: $4$, $6$, $8$, and \SI{10}{\centi\meter}. In addition, we used the realistic human head model Billie \cite{VirtualFamily},  which was also discretized with \SI{5}{\milli\meter} voxel isotropic resolution. 

\subsubsection{Tissue mimicking phantoms}\label{sec:phantoms_experimental}

\begin{figure}[t!]
\begin{center}
\includegraphics[width=0.23\textwidth]{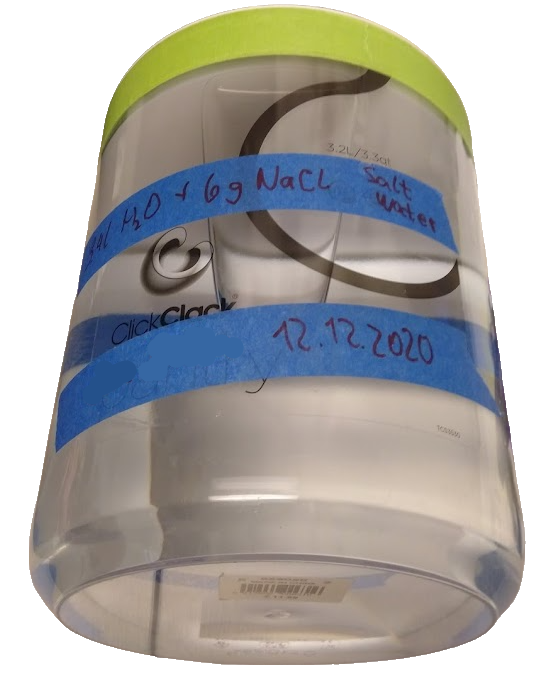}
\includegraphics[width=0.23\textwidth]{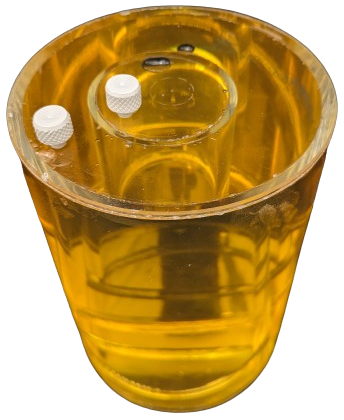}
\caption{The homogeneous phantom used for calibration (left) and two\hyp compartment phantom cylindrical phantom (right). The homogeneous phantom was filled with a solution of distilled water and sodium chloride, while for the two\hyp compartment phantom, we also added polyvinylpyrrolidone and manganese(II) chloride to control the permittivity and T1, respectively.}
\label{fig:n_phantoms}
\end{center}
\end{figure}

We constructed a single- and a two\hyp compartment cylindrical phantoms filled with tissue\hyp mimicking solutions (Fig.\,\ref{fig:n_phantoms}). {\color{black}We chose cylinders over other geometries, such as spheres, as they provide better loading for our coil (see Section \ref{sec:coil} for details)}. For the single\hyp compartment phantom, the solution contained distilled water and sodium chloride, and was placed inside an acrylic container tube with a wall thickness of \SI{0.125}{\centi\meter}, length of \SI{21}{\centi\meter} and radius of \SI{8.5}{\centi\meter}. The EP were measured with a dielectric probe (85070E Agilent, Santa Clara, CA) and were found to be $79.4$ and \SI{0.74}{\siemens\per\meter} for the relative permittivity and conductivity, respectively, at room temperature. The two\hyp compartment phantom was \SI{21.5}{\centi\meter} long with an outer radius of \SI{6.875}{\centi\meter}. The inner compartment was also a cylinder of the same length with a radius of \SI{3.1}{\centi\meter} and it was displaced from the center by roughly \SI{0.5}{\centi\meter}. The thickness of the outer compartment's acrylic container was \SI{0.5}{\centi\meter}, while the thickness of the inner container was \SI{0.3}{\centi\meter}. For mixing the solutions, we followed the recipes in \cite{ianniello2018synthesized} using distilled water, sodium chloride, polyvinylpyrrolidone, and manganese(II) chloride (probe\hyp measured values: Outer compartment: $\epsilon_r = 59$, $\sigma_e = \SI{0.7}{\siemens\per\meter}$, Inner compartment: $\epsilon_r = 75$, $\sigma_e = \SI{1.87}{\siemens\per\meter}$). The uncertainty of the probe is reported equal to $5\%$ based on the user's manual.   

\subsection{Transceiver coil}\label{sec:coil}

\subsubsection{Coil design}

\begin{figure}[t!]
\begin{center}
\includegraphics[width=0.46\textwidth]{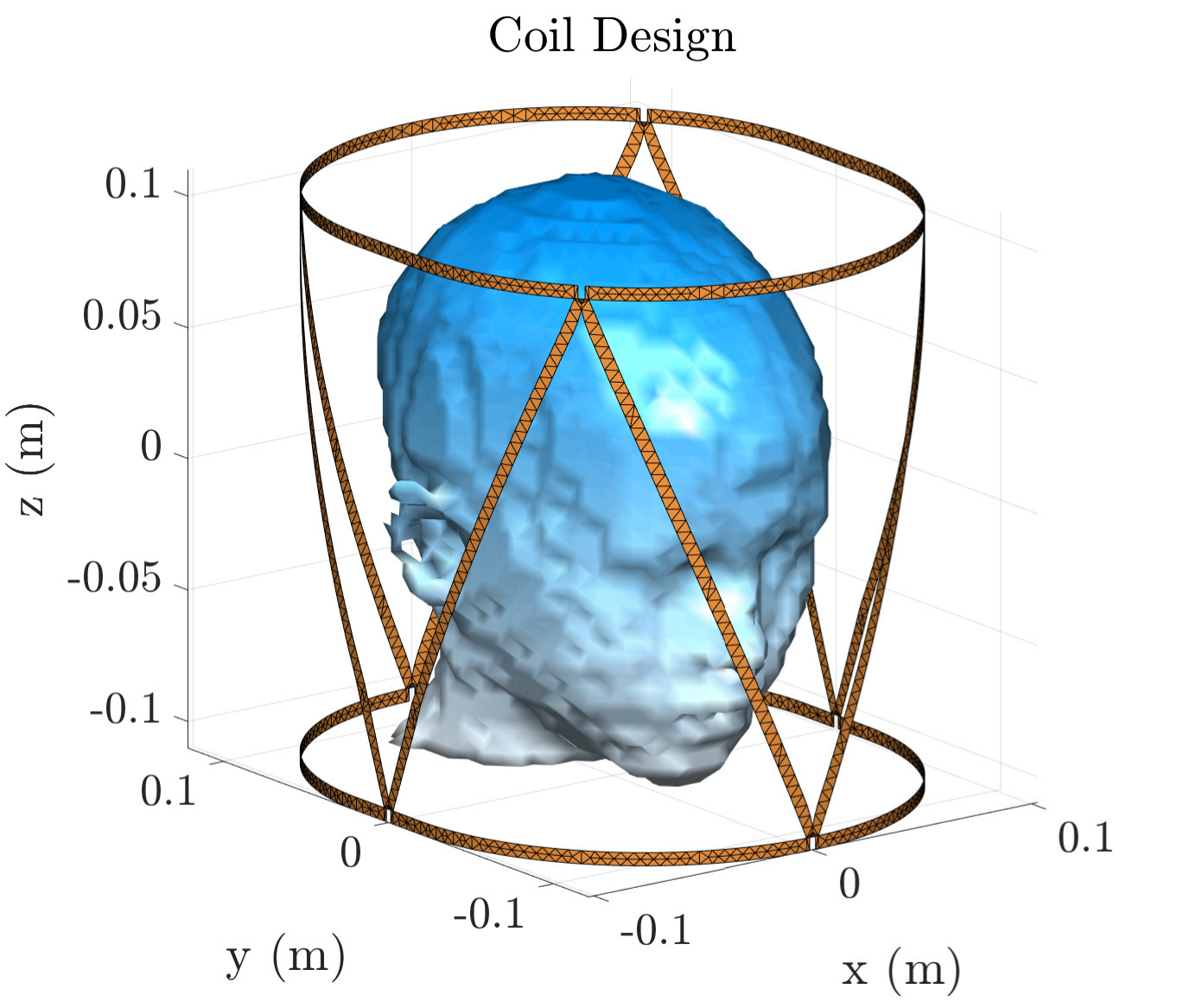}\\
\includegraphics[width=0.23\textwidth]{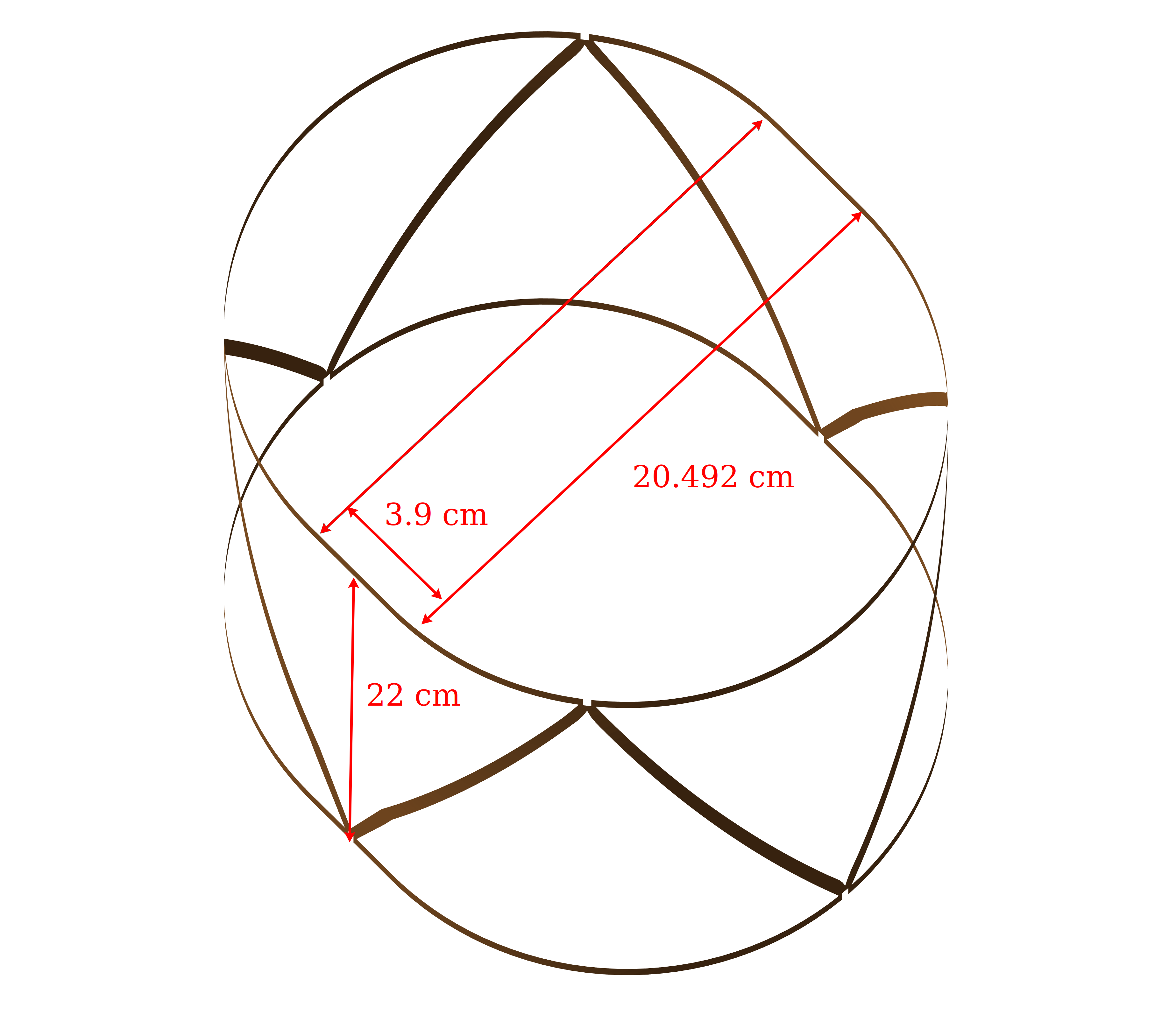}
\includegraphics[width=0.23\textwidth]{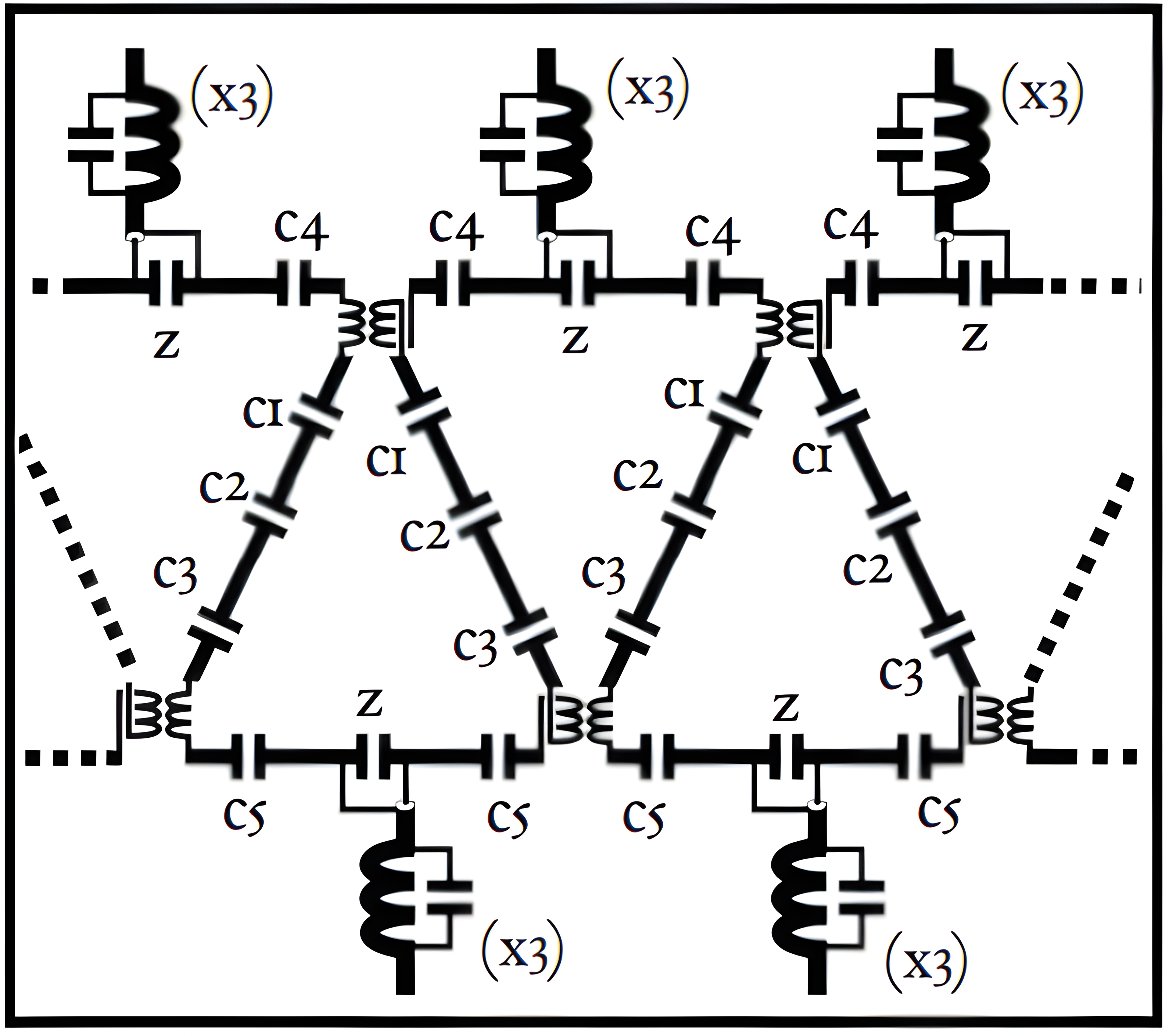}
\caption{(top) The model of the triangular RF coil array loaded with the Billie head model. {\color{black}(bottom left) Top view of the coil geometry. The annotations indicate the width of the rectangular section, the radius of the semicircles, and the overall length of the coil}. (bottom right) Circuit diagram of three representative elements of the triangular coil array. The coil array consisted of eight triangular elements that were capacitively decoupled from their nearest neighbors and inductively decoupled from their next nearest neighbors.}
\label{fig:n1}
\end{center}
\end{figure}

Our coil was inspired by designs previously proposed to introduce incident field variation in the longitudinal direction, which are beneficial in GMT \cite{chen2018highly}. We designed a \SI{7}{\tesla} head coil with eight transceiver triangular elements arranged on a stadium substrate. The length of the array is \SI{22}{\centi\meter}, the radius of the semicircles is \SI{10.246}{\centi\meter}, the width of the rectangle is \SI{3.9}{\centi\meter}, and the width of the conductors is \SI{1}{\centi\meter} (Fig.\,\ref{fig:n1}) The coil was discretized with $2140$ mesh triangles. 
\par
A circuit diagram for three array elements is shown in Fig.\,\ref{fig:n1}. Each triangular element is segmented with eight capacitors and one feeding\hyp port. The capacitors $c_1$, $c_2$, and $c_3$ were adjusted for tuning and to ensure decoupling between nearest neighbors. $c_4$ and $c_5$ were distributed on the end rings and their value was adjusted during tuning. A pair of counter\hyp wounded inductors was used for the next nearest neighbor decoupling. One parallel and one series variable capacitor are used for the matching network (denoted as $z$ in Fig.\,\ref{fig:n1}). 
\par
The capacitor values were optimized for the Billie head model. We optimized over all the capacitors until the Frobenius norm of the scattering parameter matrix was minimum to ensure coil tuning, matching, and decoupling, as in \cite{giannakopoulos2020magnetic}. The inductors and mutual inductances were instead modeled as the ones in the coil prototype (see following section).

\subsubsection{Coil prototype}

\begin{figure}[t!]
\begin{center}
\includegraphics[width=0.23\textwidth]{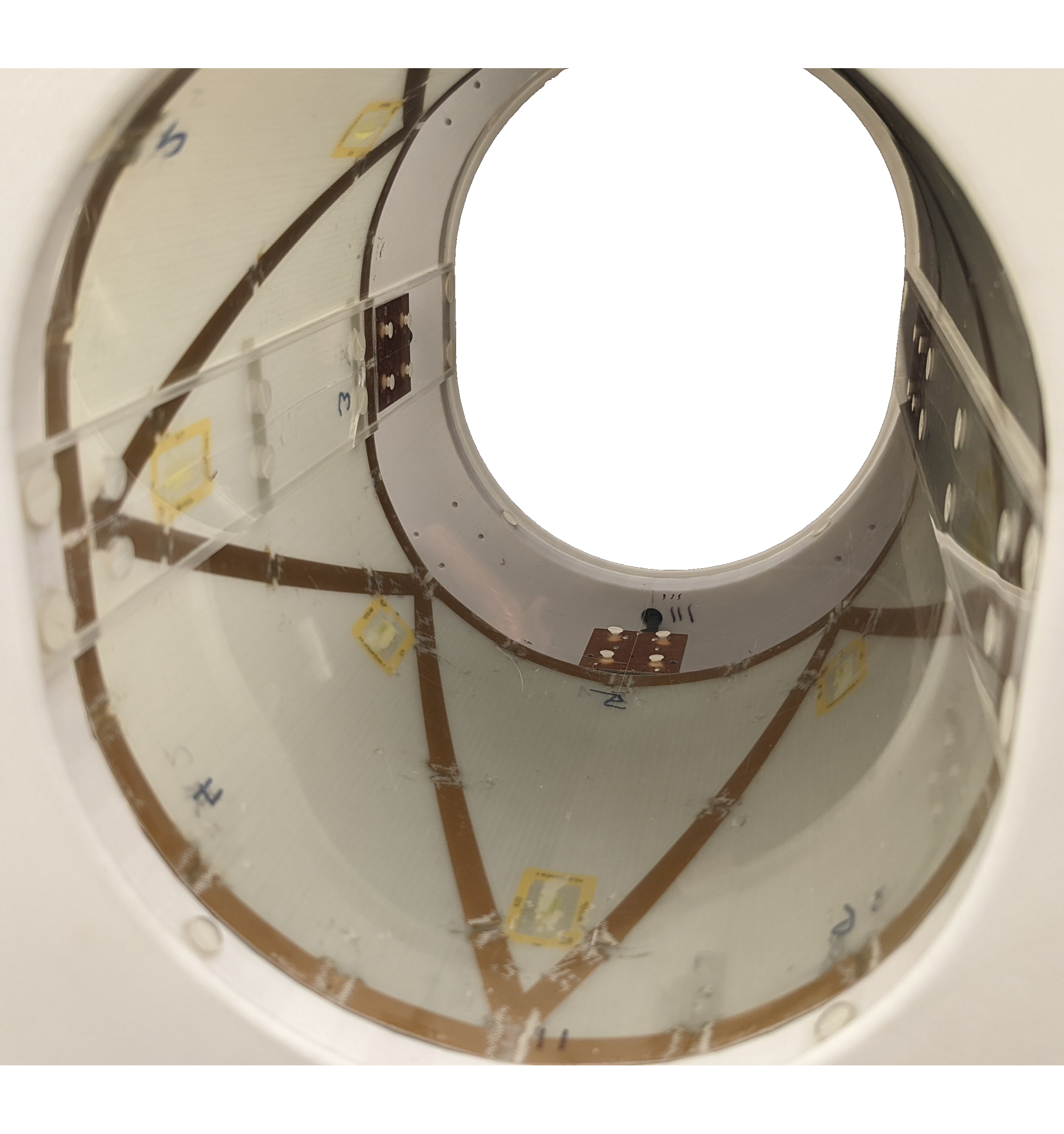} 
\includegraphics[width=0.23\textwidth]{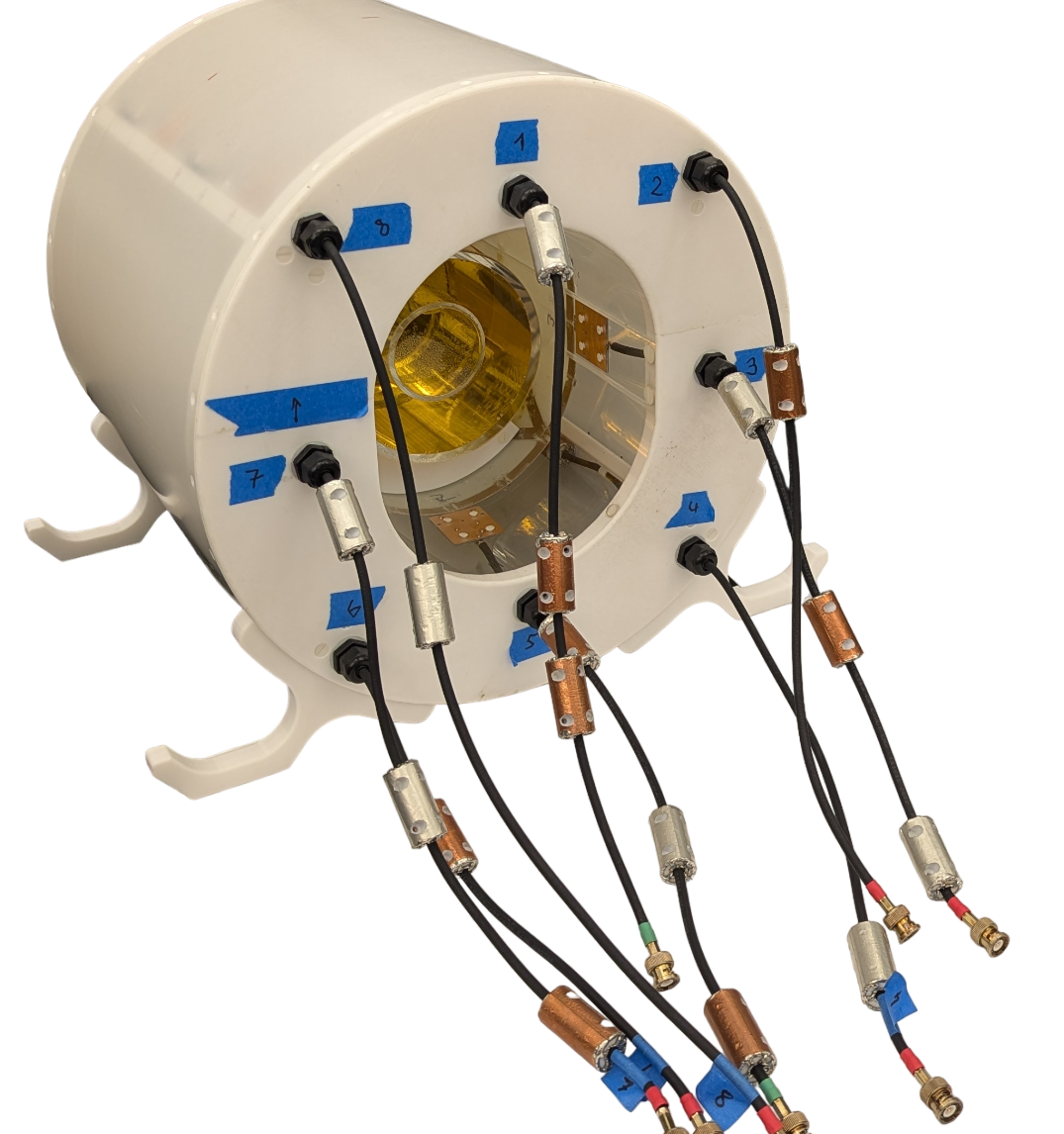}
\caption{(left) Interior of the $8$\hyp channel triangular transceiver RF coil array. Yellow skin markers were placed $6$ cm away from the end ring to help matching the location of the sample relative to the coil in the forward problem simulation. (right) The coil loaded with a cylindrical phantom.}
\label{fig:n2}
\end{center}
\end{figure}

We constructed the coil and tuned and decoupled it while loaded with the homogeneous cylindrical phantom. This resulted in $8.2$, $6.8$, $12$, $2$, and $2.2$ pF for $c_1$, $c_2$, $c_3$, $c_4$, and $c_5$, respectively. The coupled inductors were manually constructed and their self inductance varied from \SI{26}{\nano\henry} to \SI{32}{\nano\henry}. The mutual inductance coefficients varied from $-0.72$ to $0.77$ corresponding to mutual inductances between \SI{-25}{\nano\henry} and \SI{-20}{\nano\henry} nH, respectively. The array elements were tuned and decoupled so that the scattering (S) parameter matrix values were at most \SI{-16.8}{\decibel} in the diagonal elements and at most \SI{-10.1}{\decibel} dB in the off diagonal elements. We attached a Pi matching network at each feeding port to match the coil to \SI{50}{\ohm}. To characterize the effect of the matching network in the simulations, we measured the scattering (S) parameter matrix when the coil was loaded with the homogeneous cylindrical phantom using a network analyzer (Keysight Technologies, Fountain Grove, CA), and then computed a calibration matrix with the method in \cite{paska2009field}. The lumped elements configuration was also applied to the coil model and the calibration matrix was multiplied with the simulated $\matvec{B}_1^{(+)}$ maps at every GMT iteration for the reconstructions of the simulated cylinders and the phantom. Three cable traps were added to each coil cable to suppress common mode currents \cite{hernandez2020review}. Photographs of the coil and its housing are shown in Fig.\,\ref{fig:n2}.

\subsection{Simulations}

We studied the effect of the sample loading on the coil current distribution by varying the distance between the conductors and the sample. Using the transceive coil model, we calculated the synthetic $\hat{\matvec{B}}_1^{(+)}$ measurements with one forward solution of the VSIE \eqref{eq:vsie} for the four homogeneous cylinders of increasing radius (described in Section \ref{sec:phantoms}) and then performed corresponding simulated VSIE\hyp based GMT reconstructions. The regularization term was neglected since the synthetic measurements were noiseless and the weights in the GMT's data\hyp consistency term were set as in \eqref{eq:weights_sqr} (left). 
\par
We also compared the performance of our proposed VSIE\hyp based GMT against the VIE\hyp based GMT for the EP reconstruction of Billie (Fig.\,\ref{fig:n1} left). For the VIE\hyp based GMT, we computed three different incident field distributions using different guesses of the EP of Billie: (I) ground\hyp truth EP distribution, (II) mean value of the ground\hyp truth EP, (III) homogeneous initial guess of the GMT reconstruction. The synthetic $\hat{\matvec{B}}_1^{(+)}$ measurements were simulated with one forward solution of the VSIE using the ground\hyp truth EP in all cases, and then corrupted with white Gaussian noise of peak SNR $200$ as in \cite{serralles2019noninvasive}. The regularizer's weight $\alpha$ was set to $0.0002$. The weights in the data\hyp consistency term in \eqref{eq:cost_function} were set to:
\begin{equation}
w_l = \sqrt{\frac{\abs{\hat{\matvec{B}}^{(+)}_{1,l}}}{\max\!\left(\abs{\hat{\matvec{B}}^{(+)}_{1,l}}\right)}}.
\label{eq:weights_sqr}
\end{equation}

All simulated GMT reconstructions were executed for $500$ iterations starting from the same homogeneous initial guess with $\epsilon_r = 21.1$ and $\sigma_e = \SI{0.2}{\siemens\per\meter}$. 

\subsection{Experiments}

We conducted a GMT experiment with the two\hyp compartment phantom on a \SI{7}{\tesla} scanner (Magnetom, Siemens Healthineers, Erlangen, Germany). We measured $\abs{\hat{\matvec{B}}_1^{(+)}}$ maps for each coil using a magnetic resonance fingerprinting technique \cite{cloos2016multiparametric}. To mitigate artifacts, we applied a median filter to the measured maps. Additionally, we extracted the relative phases between the transmit field maps using the individual coil images. Each of the eight scans required approximately $40$ minutes, covering the entire phantom with a voxel isotropic resolution of \SI{2}{\milli\meter}. 
\par
Since coil cables are not modeled in our current VSIE formulation of the forward problem, we used the homogeneous phantom to calibrate scaling differences between simulated and experimental maps. In particular, we simulated the $\matvec{B}_1^{(+)}$ maps for the homogeneous phantom using the VSIE solver and the probe\hyp measured EP. Then, we measured the corresponding experimental maps (see above) and multiplied each one by a complex weight:  $\matvec{q}_l\cdot \hat{\matvec{B}}_1^{(+)}$, $l = 1,\dots,8$. To calculate the calibration weighs, we minimized GMT's data\hyp consistency term over the weights $\matvec{q}$ instead of the EP. The $\delta$ term in the data\hyp consistency term \eqref{eq:cost_function} took the following form: 
\begin{equation}
\matvec{\delta}_{ll'}\!\left(\matvec{q}\right) = \left(\matvec{q}_{l} \cdot \hat{\matvec{B}}^{(+)}_{1,l}\right) \circ \left(\overline{\matvec{q}_{l'}\cdot\hat{\matvec{B}}^{(+)}_{1,l'}}\right) - \matvec{B}^{(+)}_{1,l} \circ \overline{\matvec{B}^{(+)}_{1,l'}},
\label{eq:cf_exp}
\end{equation}
where Einstein summation notation is used over the indices $l'$ (first) and $l$ (second). The experimental $\hat{\matvec{B}}_1^{(+)}$ acquired for the two\hyp compartment phantom were then multiplied by the calibration weights, after scaling them by $\matvec{q} \cdot V_{2}/V_{1}$, where $V_{2} = \SI{80}{\volt}$ and $V_{1} = \SI{48}{\volt}$ were the voltages applied during the two\hyp compartment and homogeneous phantom scans, respectively. We refer to this step as \textit{cross-calibration}. We also repeated the calibration by computing the weights $\matvec{q}$ using the two\hyp compartment phantom itself and the associated EP measured with the dielectric probe (\textit{self-calibration}), in order to assess the accuracy of the \textit{cross-calibration} with the homogeneous phantom and its effect on GMT performance.
\par
The GMT reconstructions were executed for $2000$ iterations starting from the same homogeneous initial guess equal to the homogeneous phantom's EP ($\epsilon_r = 79.4$, $\sigma_e = \SI{0.74}{\siemens\per\meter}$)). To account for the discrepancy between simulated and experimental conditions, we increased the weight of the regularizer $\alpha$ from the value used in the simulations to $0.09$ and $0.11$ for the \textit{cross\hyp calibration} and \textit{self\hyp calibration} cases, respectively. In this case the weights $w$ of the GMT's data\hyp consistency term were chosen to penalize more strongly the regions of low SNR than for the simulated VSIE\hyp based GMT reconstructions: 
\begin{equation}
w_l = \frac{\abs{\hat{\matvec{B}}^{(+)}_{1,l}}}{\max\!\left(\abs{\hat{\matvec{B}}^{(+)}_{1,l}}\right)},
\label{eq:weights_nosqr}
\end{equation}
since the experimental maps are expected to be larger differences with the synthetic ones.

\subsection{Evaluation}

To evaluate the reconstructions we used the peak normalized absolute error (PNAE) following \cite{serralles2019noninvasive}. The PNAE was preferred over other error metrics, to avoid biases from voxels with small EP values. PNAE was defined as $|y-x|/\text{max}\{x\}$, where $x$ is the ground truth and $y$ is the reconstructed map. We also used the 3D structural similarity index measure (SSIM) to evaluate the quality of the experimental reconstruction.

\section{Results}\label{sec:Results}

All calculations were performed using MATLAB 9.10 and an NVIDIA V100 PCIe GPU with 32 GB of memory.

\subsection{Simulation results}

\subsubsection{Effect of the distance between coil and sample on the coil current distribution}

\begin{figure}[t!]
\begin{center}
\includegraphics[width=0.48\textwidth]{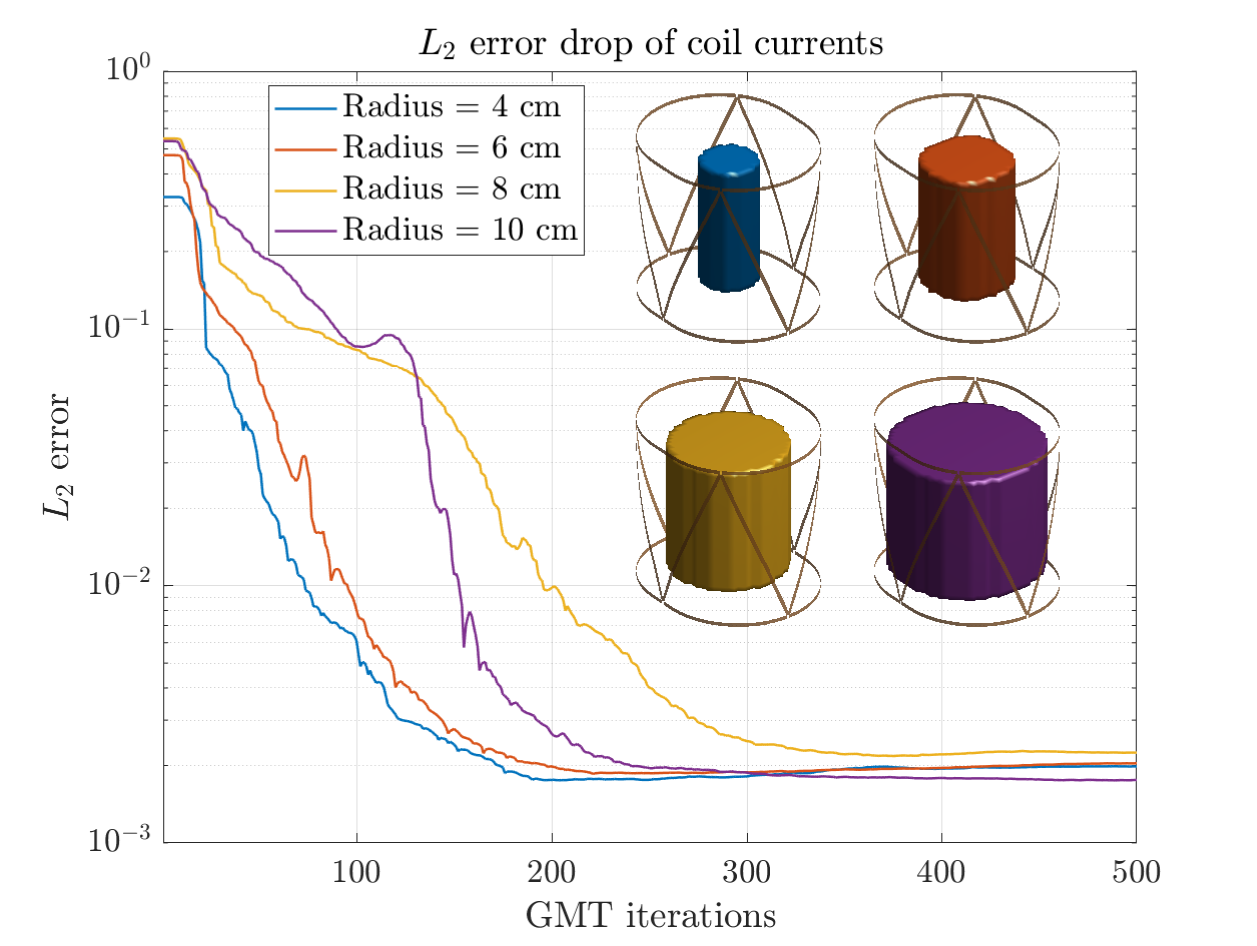}
\caption{Relative error of the $L_2$ norm of the coil currents updated in each VSIE-GMT iteration with respect to the coil currents associated with ground\hyp truth EP values. The results are presented for different cylinders to show the effect of the distance of the cylinder from the coil conductors.}
\label{fig:n5}
\end{center}
\end{figure}

Fig.\,\ref{fig:n5} illustrates the convergence of the L$_2$ relative error norm between the coil currents obtained in each GMT iteration by using the updated EP and the ground\hyp truth coil currents calculated using the actual EP values of the four homogeneous cylinders.

\subsubsection{Comparison with VIE-based GMT}

\begin{figure}[t!]
\begin{center}
\includegraphics[width=0.48\textwidth]{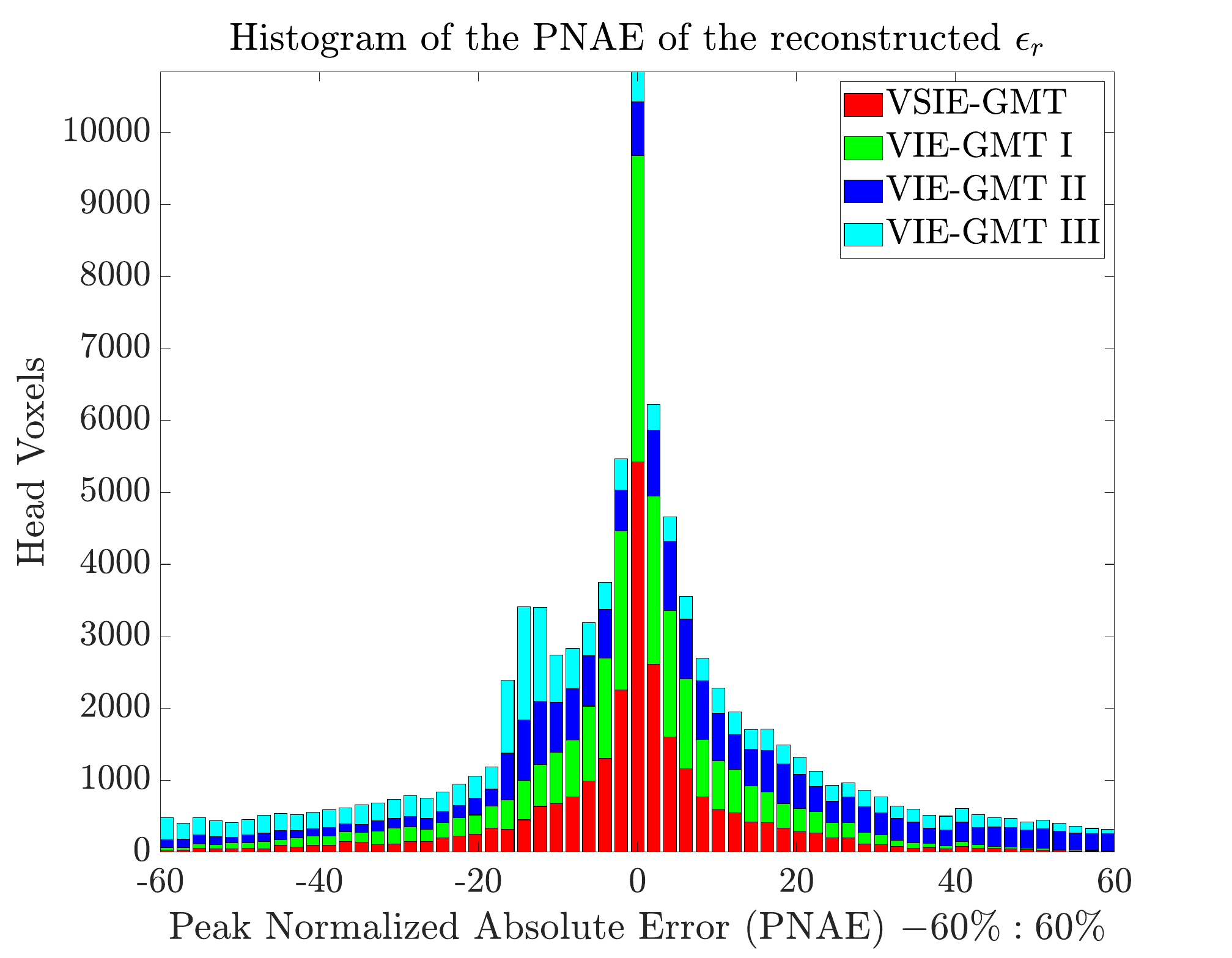}
\includegraphics[width=0.48\textwidth]{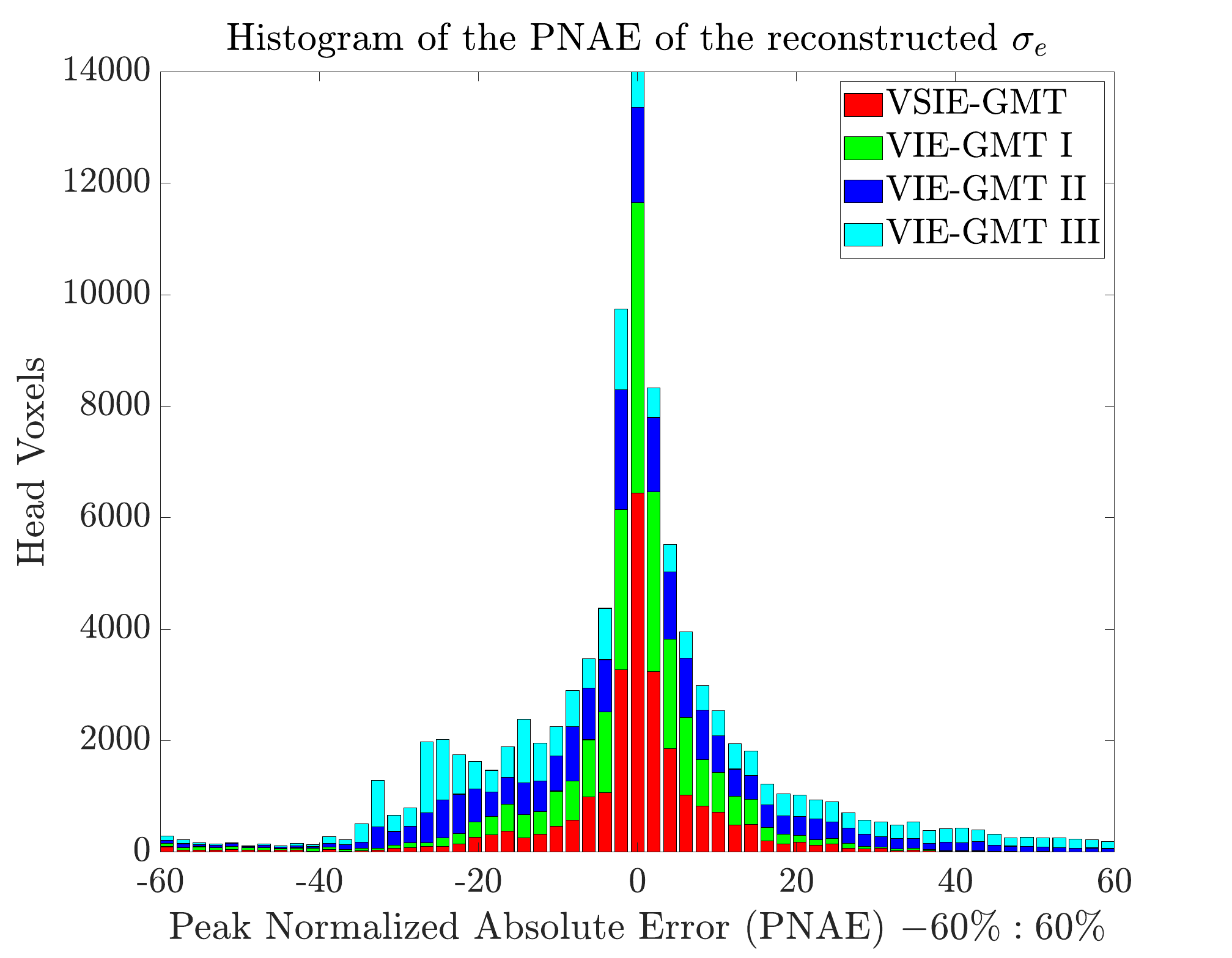}
\caption{PNAE distribution of the relative permittivity (top) and conductivity (bottom) with respect to the ground truth for all voxels in the head model and each GMT reconstruction.}
\label{fig:n6}
\end{center}
\end{figure}

\begin{figure*}[t!]
\begin{center}
\includegraphics[width=0.96\textwidth,trim={1cm 0 1cm 0},clip]{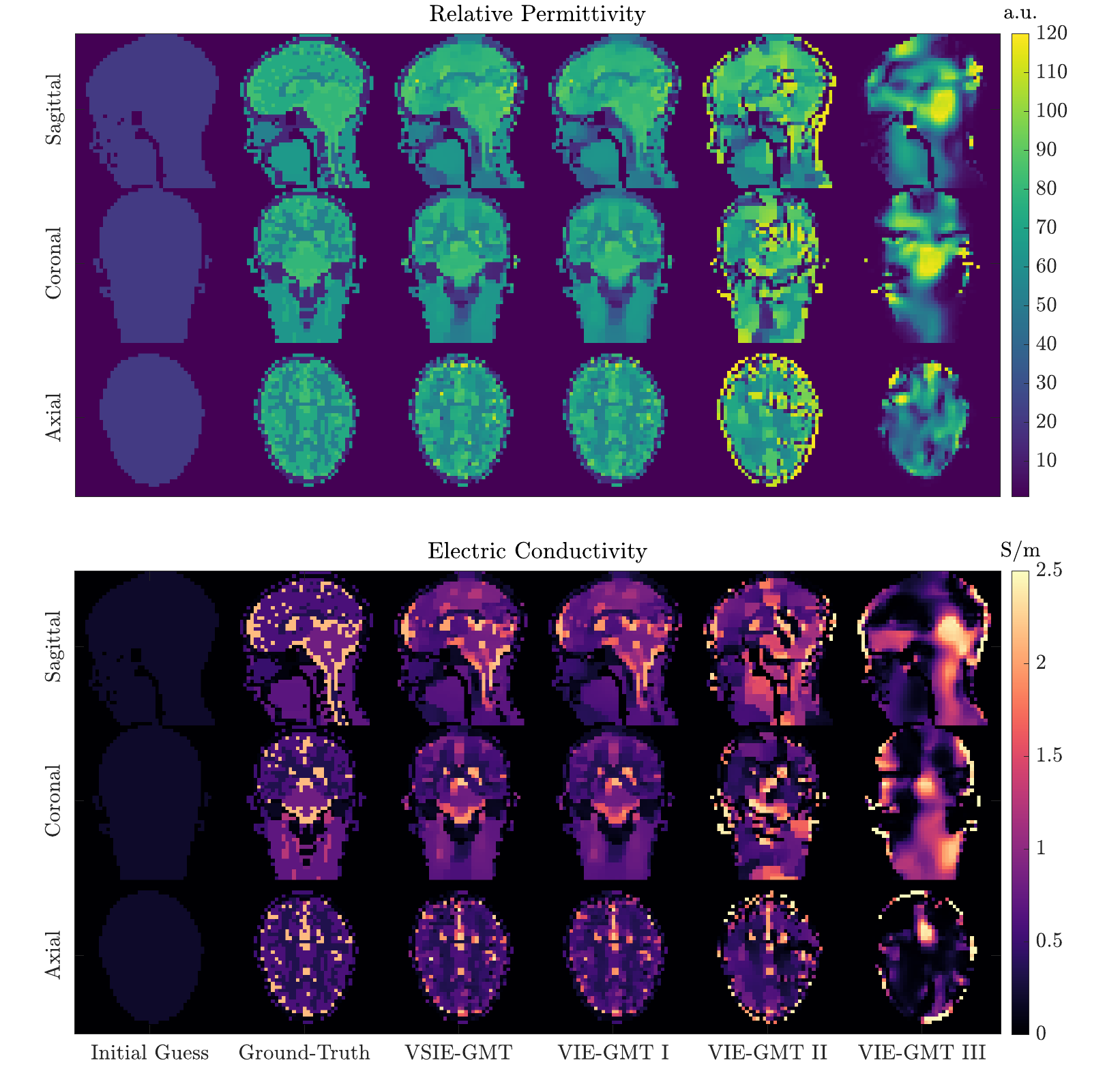} \\
\caption{Comparison between the reconstructed relative permittivity (top) and conductivity (bottom) maps {\color{black}for representative sagittal, coronal, and axial cuts} through the brain area of Billie using different EP reconstruction methods. From left to right, the initial guess, the ground\hyp truth, and the reconstructions using either the VSIE\hyp based GMT or the VIE\hyp based GMT are presented. For VIE\hyp based GMT we approximated the incident electromagnetic field using three different EP distributions for Billie: (I) ground\hyp truth EP distribution, (II) mean value of the ground\hyp truth EP, (III) homogeneous initial guess of the GMT reconstruction.}
\label{fig:n7}
\end{center}
\end{figure*}

Fig.\,\ref{fig:n6} shows stacked histograms for the four GMT reconstructions of the PNAE distribution of the relative permittivity and conductivity for all voxels in Billie's head. The mean PNAE between the ground\hyp truth EP and all the reconstructed EP is presented in Tb.\,\ref{tb:errors2}. Fig.\,\ref{fig:n7} shows a qualitative comparison between the ground\hyp truth and reconstructed EP for {\color{black}representative sagittal, coronal, and axial planes of the brain}. 

\begin{table}[ht!]
\caption{Mean PNAE between the reconstructed EP and ground\hyp truth values for Billie} \label{tb:errors2} \centering
{\def\arraystretch{1.3}\tabcolsep=6.5pt
\begin{tabular}{ c c c }
\hline
Reconstruction               & $\epsilon_r$  & $\sigma_e$ \\
\hline
VSIE\hyp based GMT           & $9.4\%$       & $7.5\%$    \\
VIE\hyp based GMT - case I   & $10.5\%$      & $8.2\%$    \\
VIE\hyp based GMT - case II  & $30.9\%$      & $19.6\%$   \\
VIE\hyp based GMT - case III & $37.9\%$      & $28.9\%$   \\
\hline						
\end{tabular}
}
\end{table}

\subsection{Experimental results}

\begin{figure}[t!]
\begin{center}
\includegraphics[width=0.48\textwidth,trim={{0.09\textwidth} {0.08\textwidth} {0.09\textwidth} {0.06\textwidth}},clip]{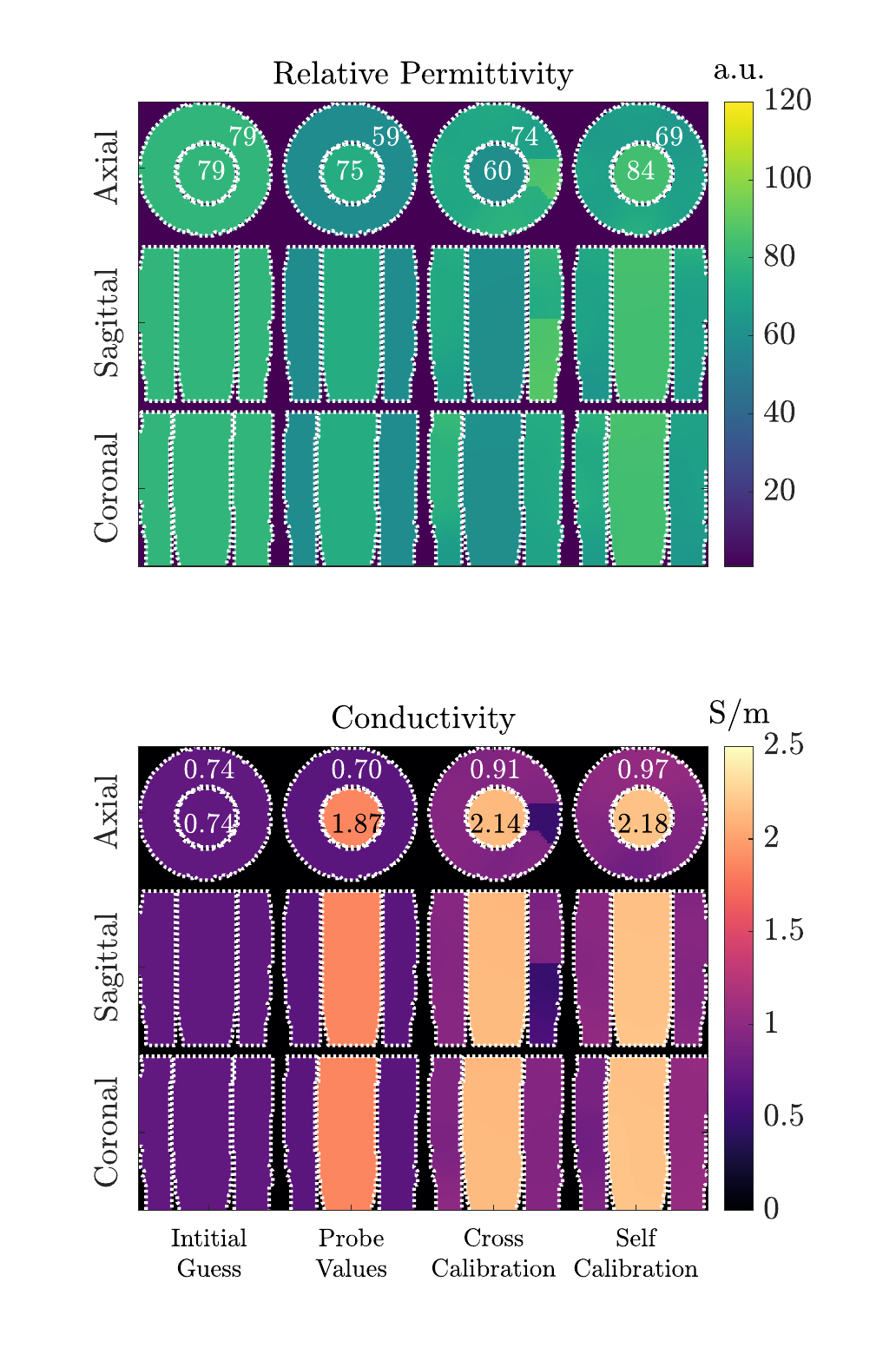}
\caption{Experimental VSIE\hyp based GMT reconstructions for the \textit{cross\hyp calibration} approach with the homogeneous phantom and the \textit{self\hyp calibration} approach. EP maps are shown for the central axial, sagittal, and coronal slices. The numbers in the axial view indicate the mean EP values in each compartment. The white dotted contours indicate the surface of the phantom and the boundary between the two compartments. The geometrical deformation at the edges of the phantom was due to the $B_0$ inhomogeneity.}
\label{fig:n8b}
\end{center}
\end{figure}

\begin{figure*}[t!]
\begin{center}
\includegraphics[width=0.96\textwidth]{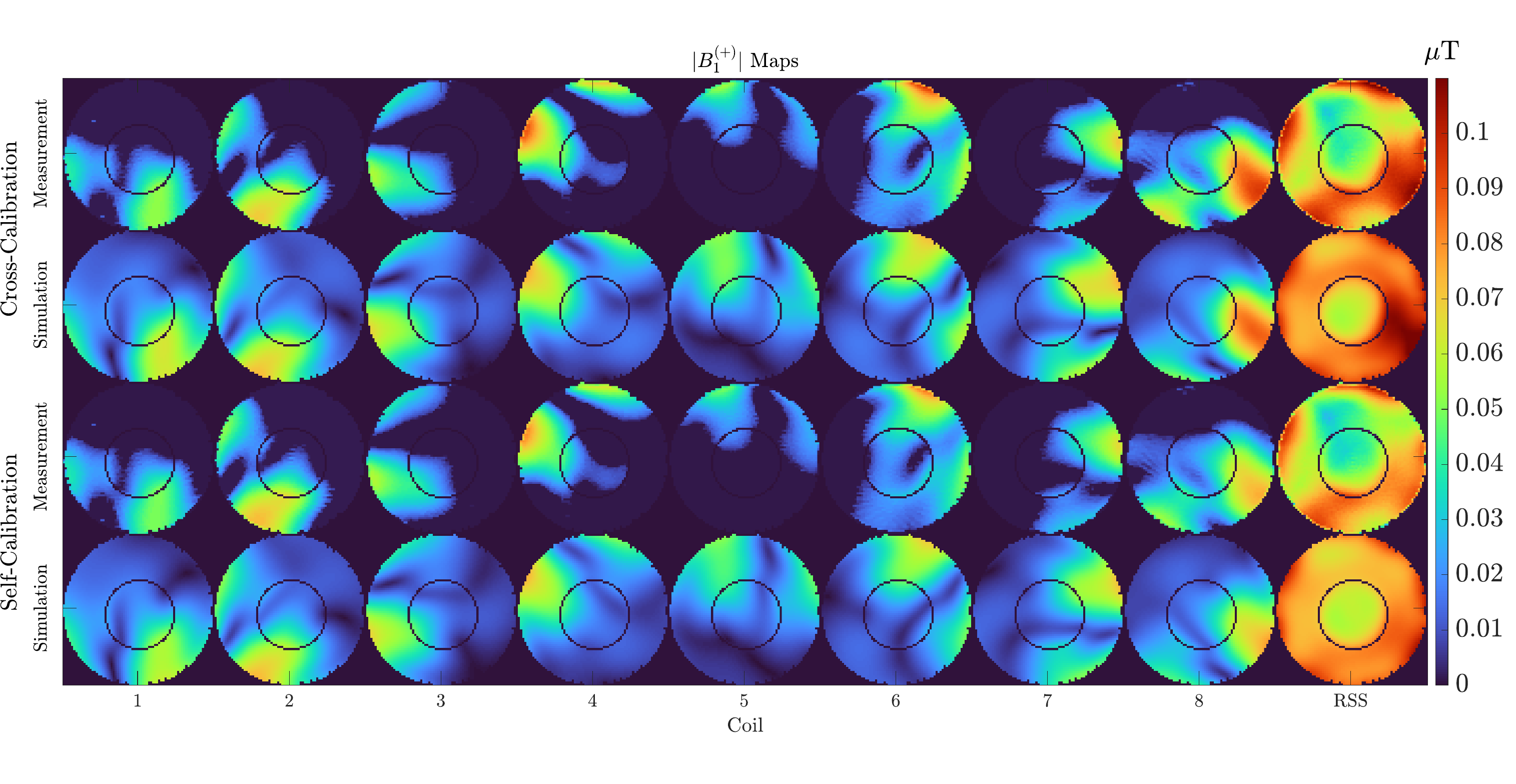}
\includegraphics[width=0.96\textwidth]{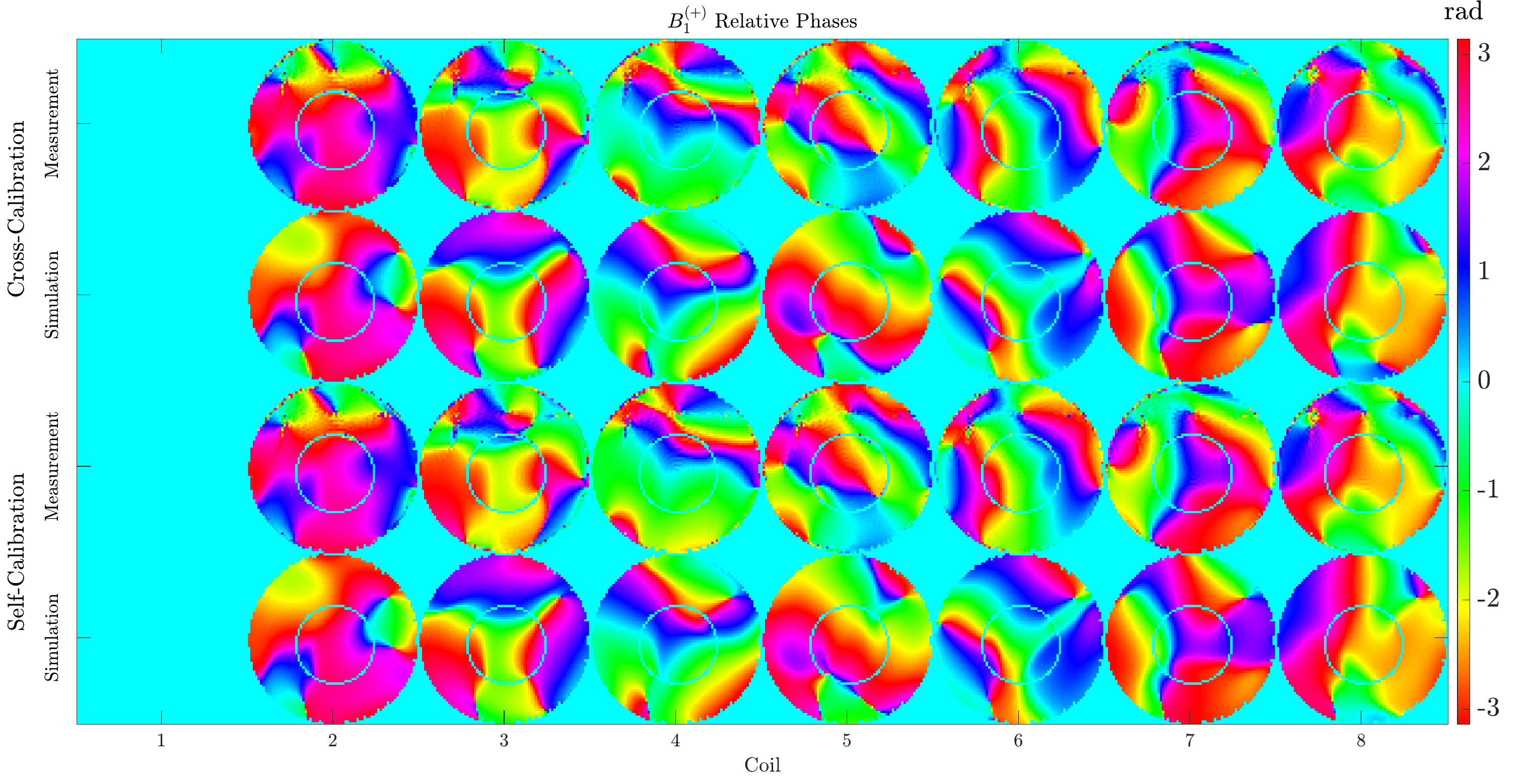} 
\caption{Measured vs. simulated fields for the central axial slice of the phantom. The measurements were performed with magnetic resonance fingerprinting and scaled using either \textit{cross\hyp calibration} or \textit{self\hyp calibration}. The simulations were performed with one forward solution of the VSIE using the reconstructed EP in each case. The $\abs{\hat{\matvec{B}}_1^{(+)}}$ patterns (top panel) were similar between measurements and experiments. The phases of $\hat{\matvec{B}}_1^{(+)}$ relative to the first coil (bottom panel) {\color{black}were also similar for most of the coils}. The rightmost column in the magnitude plots shows the root sum of squares of the maps to show that there were no regions where the field was zero for all coils.}
\label{fig:n4}
\end{center}
\end{figure*}

Fig.\,\ref{fig:n8b} shows the initial guess of EP, the probe\hyp measured EP, and the reconstructed EP with both calibrations approaches. The results are shown for the central planes of the phantom. The SSIM and the mean PNAE over all voxels of the phantom with respect to the dielectric probe values are reported in Table\,\ref{tb:errors}. Fig.\,\ref{fig:n4} compares the magnitude and the relative phases of the experimentally measured $\hat{\matvec{B}}_1^{(+)}$ maps with the corresponding maps simulated using the EP estimated by VSIE\hyp based GMT. The experimental maps were scaled using either the \textit{cross\hyp calibration} or the \textit{self\hyp calibration} approach. 

\begin{table}[ht!]
\caption{Accuracy of VSIE\hyp GMT with respect to probe-measured EP} \label{tb:errors} \centering
{\def\arraystretch{1.3}\tabcolsep=6.5pt
\begin{tabular}{ c c c c }
\hline
Metric                & EP                    & \textit{cross\hyp calibration} & \textit{self\hyp calibration}  \\
\hline
\multirow{2}{*}{PNAE} & $\epsilon_r$          & $20\%$                         & $13\%$                         \\
                      & $\sigma_e$            & $13\%$                         & $15\%$                         \\
\multirow{2}{*}{SSIM} & $\epsilon_r$          & $0.75$                         & $0.78$                         \\
                      & $\sigma_e$            & $0.93$                         & $0.93$                         \\
\hline						
\end{tabular}
}
\end{table}

\section{Discussion}\label{sec:Discussion}

In VIE\hyp based GMT \cite{serralles2019noninvasive}, the incident fields from the coils to the sample were calculated once with a guess of EP and used throughout the GMT optimization. In the first experimental proof-of-principle of GMT, the incident fields were calculated for a uniform phantom using ground\hyp truth EP measured with a dielectric probe \cite{serralles2019noninvasive}. This was equivalent to assuming that the incident fields were both known and not affected by the EP updates at each GMT iteration. Although the former was not realistic and was only adopted to demonstrate the GMT algorithm, the latter is a common assumption for EP reconstruction techniques based on the integral form of Maxwell's equation, such as CSI\hyp EPT \cite{balidemaj2015csi}, which rely on using a constant approximation of the incident fields. In this work, we integrated the VSIE formulation into GMT, eliminating the need for a separate initial simulation step to estimate the incident fields. In the VSIE\hyp based GMT, the incident EM fields are instead implicitly re\hyp estimated during each GMT iteration, accounting for the effect of the updated EP on the coil currents.
\par
There are several methods for fast solutions of the VSIE. Besides fully assembling $Z_{\rm cb}$ in \eqref{eq:vsie}, one can employ the magnetic resonance Green's function (MRGF) \cite{Villena2016} for rapid VSIE solutions. However, this approach is most effective when one needs to simulate various coil designs loaded with the same object, but it is not efficient for multiple simulations of the same coil with varying object's EP, as in GMT. Alternative methods, such as those based on tensor decompositions (ACA+Tucker \cite{giannakopoulos2021compression}, TT\hyp SVD \cite{giannakopoulos2021atensor}, Perturbation Matrix \cite{guryev2022marie}), are typically preferred when $Z_{\rm cb}$ exhibits a good low\hyp rank approximation. However, this is not the case when the coil closely fits the object, as in the simulations presented in this work. Therefore, we utilized the precorrected Fast Fourier Transform method (pFFT) method \cite{guryev2019fast, phillips1997precorrected}, which transforms the VSIE system into an equivalent VIE system by extending the volumetric computational grid to enclose both the coil and the object loading it. {\color{black}The system can then be solved rapidly by utilizing FFT \cite{Polimeridis2014}, and Tucker decomposition\hyp based acceleration methods \cite{giannakopoulos2019memory}. Since our coil closely fits the object, the extended volumetric grid exactly matches the volumetric grid used in the VIE\hyp based GMT, ensuring that our proposed VSIE\hyp based GMT can be executed as quickly as the VIE\hyp based GMT, requiring roughly two days ($500$ iterations) to reconstruct EP for a sample discretized at \SI{5}{\milli\meter}) voxel isotropic resolution, and roughly ten days ($2000$ iterations) to reconstruct EP for a sample discretized at \SI{2}{\milli\meter} voxel isotropic resolution}. Other integral equation formulations could also be employed for the solution of the VSIE \cite{eibert2009surface, Oijala2014}. Note however that a different formulation or method would require different calculations for the gradient (or the Hessian \cite{serrallesapplication} if a full Newton optimization algorithm is used) of the data\hyp consistency term, since the VSIE linear system has to be adjusted.
\par
We show that the coil currents and, consequently, the incident fields, can differ between $30\%$ and $50\%$ from their expected ground\hyp truth value (Fig.\,\ref{fig:n5}), depending on the distance between the sample and the coil, when they are estimated using the same initial guess for the sample's EP ($\epsilon_r = 21.1$, $\sigma_e = \SI{0.2}{\siemens\per\meter}$). The VSIE\hyp based GMT optimization re-calculated the coil currents for each update of the EP, which resulted in the error decreasing to $10\%$ after iterations $23$, $38$, $75$, and $90$ for the cylinders with radii of \SI{4}{\centi\meter}, \SI{6}{\centi\meter}, \SI{8}{\centi\meter}, and \SI{10}{\centi\meter}, respectively. The above show that the closer the sample is to the coil's conductors, the greater the perturbations in the currents due to changes in the EP. Therefore, the assumption of constant incident fields is valid only in the case of remote transmit coils, such as the scanner body coil at \SI{3}{\tesla}. For experimental setups involving human subjects, this limitation can become critical, as coil loading varies significantly with subject positioning, thereby introducing substantial subject-specific changes in coil currents. As a result, failing to accurately estimate these currents can severely compromise the reliability of EP reconstructions obtained through integral equation \hyp based EP reconstructions.
\par
Even when using incident fields based on the true EP distribution, the VIE\hyp GMT yielded slightly worse performance than the VSIE\hyp based GMT (VIE\hyp GMT I vs. VSIE\hyp GMT in Fig.\,\ref{fig:n7}). When the estimate of the incident fields was not based on the true EP distribution, which would be the case in actual experiments, the error in the reconstructed EP was considerably higher (VIE\hyp GMT II and III). The complete breakdown of the technique for case III occurs because the inverse problem of GMT is ill\hyp posed. This is due to having to estimate EP from incomplete measurements of the magnetic fields (i.e., the $\hat{z}$ component of the magnetic field and the absolute phase of the $\hat{\matvec{B}}_1^{(+)}$ cannot be measured in MRI), while the EP mainly affect the electric fields, for which we cannot obtained tomographic maps. As a result, small inaccuracies in the incident fields can lead to large errors in the estimated EP distribution so the uniqueness theorem will be satisfied \cite{balanis2012advanced} (i.e., one set of magnetic field measurements corresponds to a unique set of EP). The proposed VSIE\hyp based GMT formulation is more robust to these errors, as it does not require prior knowledge of the incident electromagnetic fields and instead relies only on a close match between the simulated coil model and the actual prototype. Alternatively, one could optimize the parameters of the regularizer in GMT's cost function \eqref{eq:cost_function} to try to improve the reconstructions for the VIE\hyp based GMT in the case of inaccurate estimates of the incident fields. Nevertheless, this could yield a prohibitively high computational cost because multiple GMT reconstructions would have to be performed to find the optimal parameters.
\par
The VSIE-based GMT accurately reconstructs both relative permittivity and conductivity across the entire volume of a realistic head model in simulation, without relying on the assumptions inherent in other methods \cite{van2012b}. Techniques based on the partial differential equation form of Maxwell's equations are typically two-dimensional and prone to noise amplification during reconstruction. To mitigate this, various strategies have been employed, including reformulating the problem as a convection-reaction equation (cr\hyp MREPT)\cite{hafalir2014convection}, applying Savitzky\hyp Golay filters \cite{savitzky1964smoothing} for computing numerical derivatives of $\matvec{B}_1^{(+)}$ \cite{liu2015gradient}, incorporating induced current images (first\hyp order current density EPT) to improve the reconstruction \cite{fuchs2018first}, or using the Green's integral identity (Green EPT) \cite{zilberti2025magnetic}. While these approaches are computationally faster than integral methods like the one presented here, they introduce artifacts at tissue boundaries and sometimes produce noisy or blurred reconstructions \cite{hafalir2014convection, balidemaj2015feasibility, eda2021method}. 
\par 
The VSIE\hyp based GMT achieved comparable accuracy as the VIE\hyp based GMT \cite{serralles2019noninvasive} in experiments for a homogeneous phantom \cite{giannakopoulosnovel}, with both approaches yielding less than $10\%$ relative error for both electrical properties. Note that the VIE\hyp GMT relied on an $8 \times 8$ calibration weight matrix, whereas VSIE\hyp based GMT uses only $8$ weights to calibrate the experimental measurements. The weight matrix effectively calibrated the field from each channel as a linear combination of the fields of all channels. While this approach was flexible, it could introduce potential ambiguities in the relative phases by allowing channel crosstalk to be absorbed into the calibration, which in turn can affect reconstruction accuracy. For inhomogeneous phantoms, VSIE\hyp based GMT achieved greater per-compartment homogeneity compared to previously proposed techniques. For example, the standard deviation of the conductivity within the inner compartment in Fig.~\ref{fig:n8b} was below \SI{0.12}{\siemens\per\meter}, whereas for std\hyp MREPT and Voigt's method (Fig.~17(a,b) in \cite{hafalir2014convection}) the reconstructed EP were highly inhomogeneous inside a uniform compartment of a two\hyp compartment phantom, with conductivity values ranging from approximately $0.4$ to \SI{1.4}{\siemens\per\meter}, despite a probe\hyp measured ground truth conductivity of approximately \SI{1.05}{\siemens\per\meter}. For cr\hyp MREPT and constrained cr\hyp MREPT (Fig.~17(c,d) in \cite{hafalir2014convection}) the reported range of conductivity values reduces to approximately $0.7$ to \SI{1.0}{\siemens\per\meter} inside the same phantom compartment. Nevertheless, note that these results can only be compared qualitatively, not quantitatively, since our experimental setup differs from previous studies in phantom design, resolution, coil geometry, and field strength. Finally, learning-based methods offer even greater computational speed and have shown great promise in in vivo reconstructions \cite{mandija2019opening, hampe2020investigating, giannakopoulos2024mr}, but their reliability remains unclear, making them difficult to fully trust for clinical translation yet. {\color{black}Approaches that combine physics with deep learning, such as \cite{guo2023three, guo2023physics}, offer a promising alternative direction by enforcing the governing partial differential equations during training. In the context of EP reconstruction, methods like PIFON-EPT  \cite{yu2023pifon}) can simultaneously learn the EP distribution and denoise (and/or complete) the measured $\hat{\matvec{B}}_1^{(+)}$ field at any location within the object. However, because these methods typically employ relatively small, fully connected networks rather than convolutional architectures, they lack the expressive power needed to accurately reconstruct highly inhomogeneous objects.} 
\par
To account for differences between the simulated and experimental coil setup, we performed a calibration of the $\hat{\matvec{B}}_1^{(+)}$ using a separate homogeneous phantom for which the EP were measured with a dielectric probe. We repeated the calibration also with the inhomogeneous phantom itself, using the probe-measured EP. This is an unrealistic scenario since it assumes that one already knows the EP they want to measure, but we performed it to verify whether calibration inaccuracy could partially explain the observed error between reconstructed and probe\hyp measured EP. However, even when using such \textit{self\hyp calibration} approach, the PNAE did not change considerably. In particular, the PNAE for the inner (outer) compartment was $13\%$ ($17\%$) and $17\%$ ($39\%$) for the permittivity and conductivity, respectively, compared to $13\%$ ($26\%$) and $17\% (33\%)$ achieved with the \textit{cross\hyp calibration}. In the case of in vivo reconstructions, the role of the homogeneous phantom in the \textit{cross\hyp calibration} could potentially be fulfilled by registering the EP of a realistic head model \cite{VirtualFamily} to the subject’s 3D MRI.
\par
Although the measurements included regions of low signal, these were weighted low (see \eqref{eq:weights_nosqr}) during the GMT reconstructions and Fig.\,\ref{fig:n4} shows that overall the magnitude maps had similar field distributions between the calibrated measurements and the simulations. This suggests that the calibration was relatively accurate in both cases and there are other discrepancies between the experimental and simulation setups that contributed to the errors in the reconstructed EP. These could include radiation losses from the cables feeding the coil, which are not considered in the VSIE simulations. Another potential source of error not modeled in the VSIE framework is external electromagnetic radiation which can potentially be mitigated by using a shielded coil for data acquisition \cite{wald2005design}. In this case, one needs to include the shield in the simulation and solve a hybrid VSIE as in \cite{giannakopoulos2022hybrid}. Furthermore, the measured $\hat{\matvec{B}}_1^{(+)}$ may be affected by spurious phase offsets (Fig.\,\ref{fig:n4}), further contributing to errors in EP estimation \cite{gavazzi2019accuracy}. One approach to eliminate these error sources would be to shield the coil and further improve the model of the coil setup in the VSIE simulation. This would eventually reduce the error between simulated and experimental measurements. {\color{black}Nevertheless, it remains challenging to fully capture these residual errors in the incident fields using the VSIE formulation alone. Moreover, VSIE\hyp based GMT relies on accurate knowledge of the coil, which is a limitation, because it is difficult to consistently ensure a perfect match between the simulated coil design and the experimental prototype. To address this, future work, will explore approximating residual differences between simulated fields and experimental measurements using a weighted combination of modes from a numerical electromagnetic basis \cite{georgakis2022novel}. This approach will require jointly solving for both the weights and the EP, which can introduce a significant increase in the computational complexity of the inverse problem, as described in \cite{serralles2023simultaneous}.}

\section{Conclusion}\label{sec:Conclusion}

This work introduced a new formulation of GMT that utilizes the VSIE to account for perturbations in the coil currents due to the updated EP of the sample in each iteration of the optimization. We demonstrated VSIE\hyp based GMT in simulations with a realistic head model. We built a \SI{7}{\tesla} transceiver coil and validated the technique also in a phantom experiment. Future work will focus on reducing the error between experimental and simulated measurements, optimizing GMT's convergence, and reducing acquisition speed to enable in vivo experiments.

\subsection*{Conflict of Interest Statement}\label{sec:CIS}

JECS and RL have a patent related to the topic of this manuscript (Dkt. No: 046434-0601).

\end{document}